\documentclass[pra,reprint,numerical,superscriptaddress]{revtex4-1}
\usepackage{graphicx}
\usepackage{dcolumn}
\usepackage{bm}

\usepackage[utf8]{inputenc}
\usepackage[T1]{fontenc}
\usepackage{mathptmx}
\usepackage{etoolbox}
\usepackage{appendix}
\usepackage{natbib}
\usepackage{amsmath}
\usepackage{braket}
\usepackage{xcolor}
\usepackage{mathtools}
\usepackage{soul}
\usepackage{subcaption}
\usepackage[font=small,labelfont=bf,justification=justified,format=plain]{caption}

\usepackage[compact]{titlesec}
    \titlespacing{\section}{1pt}{2ex}{1ex}
    \titlespacing{\subsection}{0pt}{1ex}{0ex}
    \titlespacing{\subsubsection}{0pt}{0.5ex}{0ex}
    
\newcommand{\phizero}{\Phi_0^2}
\newcommand{\phizerotwo}{\frac{\phizero}{2}}
\newcommand{\vari}[1]{\varphi_{#1}}

\renewcommand{\thefigure}{\Roman{figure}}

\newcommand{\qd}{q^{\dag}}
\newcommand{\cd}{c^{\dag}}

\makeatletter
\def\@email#1#2{%
 \endgroup
 \patchcmd{\titleblock@produce}
  {\frontmatter@RRAPformat}
  {\frontmatter@RRAPformat{\produce@RRAP{*#1\href{mailto:#2}{#2}}}\frontmatter@RRAPformat}
  {}{}
}%

\makeatother
\begin{document}


\title[Single Shot i-Toffoli Gate in Dispersively Coupled Superconducting Qubits ]{Single Shot i-Toffoli Gate in Dispersively Coupled Superconducting Qubits }
\author{Aneirin J. Baker}
\affiliation{SUPA, Institute of Photonics and Quantum Sciences,
Heriot-Watt University, Edinburgh EH14 4AS, United Kingdom}
\email{ajb17@hw.ac.uk}

\author{Gerhard B. P. Huber}%
\affiliation{Department of Physics, Technical University of Munich, 85748 Garching, Germany}
\affiliation{Walther-Mei{\ss}ner-Institut, Bayerische Akademie der Wissenschaften, 85748 Garching, Germany}

\author{Niklas J. Glaser}%
\affiliation{Department of Physics, Technical University of Munich, 85748 Garching, Germany}
\affiliation{Walther-Mei{\ss}ner-Institut, Bayerische Akademie der Wissenschaften, 85748 Garching, Germany}

\author{Federico Roy}%
\affiliation{Walther-Mei{\ss}ner-Institut, Bayerische Akademie der Wissenschaften, 85748 Garching, Germany}
\affiliation{Theoretical Physics, Saarland University, 66123 Saarbr\"ucken, Germany}

\author{Ivan Tsitsilin}%
\affiliation{Department of Physics, Technical University of Munich, 85748 Garching, Germany}
\affiliation{Walther-Mei{\ss}ner-Institut, Bayerische Akademie der Wissenschaften, 85748 Garching, Germany}

\author{Stefan Filipp}%
\affiliation{Department of Physics, Technical University of Munich, 85748 Garching, Germany}
\affiliation{Walther-Mei{\ss}ner-Institut, Bayerische Akademie der Wissenschaften, 85748 Garching, Germany}
\affiliation{Munich Center for Quantum Science and Technology (MSQCT), Schellingstra\ss e 4, 80799 M\"{u}nchen, Germany}

\author{Michael J. Hartmann}
\affiliation{Friedrich-Alexander University Erlangen-N\"urnberg (FAU), Department of Physics, Erlangen, Germany and Max Planck Institute for the Science of Light, Erlangen, Germany}

\date{\today}
\begin{abstract}
Quantum algorithms often benefit from the ability to execute multi-qubit (>2) gates. To date such multi-qubit  gates are typically decomposed into single- and two-qubit gates, particularly in superconducting qubit architectures. The ability to perform multi-qubit operations in a single step could vastly improve the fidelity and execution time of many algorithms.
Here, we propose a single shot method for executing an i-Toffoli gate, a three-qubit gate gate with two control and one target qubit, using currently existing superconducting hardware. We show numerical evidence for a process fidelity over $98\%$ and a gate time of $500$ ns for superconducting qubits interacting via tunable couplers. Our method can straight forwardly be extended to implement gates with more than two control qubits at similar fidelities.
\end{abstract}
\maketitle
The implementation of gate based quantum algorithms has made ground breaking advances in recent years, particularly in superconducting circuit architectures \cite{Arute2019}. Quantum computing is thus entering the so called Noisy Intermediate Scale Quantum Computer (NISQ) era \cite{Preskill2018QuantumBeyond}, where devices are getting powerful enough to challenge classical computing power but fall short of allowing for implementation of quantum error correction. Despite this remarkable progress, achievable gate fidelities still limit the number of gates that can be executed in a circuit and thus limit applications. 

A possible step forward could be to replace multi-qubit gate decomposition's by a single multi-qubit gate. The Toffoli or controlled-controlled Not (CCX) gate, for example, requires at least six CNOT gates \cite{Markov2008OnGates} and other single qubit gates in it's decomposition. This severely effects the fidelity and gate time of such higher order gates. For a single step multi-qubit gate to be helpful, it needs to be executed with better fidelity and shorter execution time than the equivalent decomposition. Particularly in superconducting circuits, the attention has so far been focused on the latter since the qubits typically only interact with their direct neighbors.
These higher order gates are however crucial ingredients of more complex algorithms such as Quantum Error Correction \cite{Shor1995SchemeMemory}, Grover's Search Algorithm \cite{Grover1996ASearch}, and algorithms for Quantum Chemistry \cite{Cao2019QuantumComputing,McArdle2020QuantumChemistry}. The Toffoli gate (which is required for designing quantum analogues of classical algorithms) is a prime example of a higher order gate that would benefit from a single shot implementation. 

Here we propose a mechanism for performing higher order gates on current superconducting hardware and architectures, using a recent idea \cite{Rasmussen2020Single-stepGates} adapted for a more viable implementation in readily existing hardware.  We utilize the ZZ couplings that can be engineered with capacitive or tunable couplers in superconducting circuits (SCCs) via dispersive shifts in the qubit transition frequencies \cite{Xu2021ZZGates,Collodo2019ObservationResonators,Sung2021RealizationCoupler}. We note here that ZZ couplings that could generated by the nonlinear interaction originating from directly connecting the qubits via a Josephson junction \cite{Rasmussen2020Single-stepGates,Collodo2019ObservationResonators} do not lead to scalable lattices since they generate closed loops, where flux quantization makes the device highly  sensitive to flux noise.

In our system, we consider dispersive ZZ-interactions that shift the transition frequencies of the qubits conditioned on the states of the qubits that they interact with. Selecting one qubit as the target qubit, we can thus  apply a single qubit drive on this $ \ket{0} \leftrightarrow \ket{1}$ transition, which is dispersively shifted by the ZZ-interactions. This drive executes a high-fidelity flip of this target qubit, if and only if the two remaining qubits are in the $\ket{1}$ states. For our system this results in the $\ket{101} \leftrightarrow \ket{111}$ transition, where the first and third qubits are the controls and the second qubit the target qubit. All other input states remain invariant. As we show below, this scheme thus leads to the implementation of an i-Toffoli gate, which has a matrix representation
\begin{eqnarray}\label{eqt:UITOFF}
\nonumber
U_{i-Toffoli} = 
    \begin{pmatrix} 
    1 & 0 & 0 & 0 & 0 & 0 & 0 & 0 \\
    0 & 1 & 0 & 0 & 0 & 0 & 0 & 0 \\
    0 & 0 & 1 & 0 & 0 & 0 & 0 & 0 \\
    0 & 0 & 0 & 1 & 0 & 0 & 0 & 0 \\
    0 & 0 & 0 & 0 & 1 & 0 & 0 & 0 \\
    0 & 0 & 0 & 0 & 0 & 0 & 0 & -i \\
    0 & 0 & 0 & 0 & 0 & 0 & 1 & 0 \\
    0 & 0 & 0 & 0 & 0 & -i  & 0 & 0 \\
    \end{pmatrix}
\end{eqnarray}
in the computational basis.

For this scheme to work with high-fidelity, the dispersively shifted transitions need to be individually addressable, which requires the drive amplitude to be smaller than the shifts\cite{Khazali2020FastCircuits}. The gate time is therefore only dependent on the strength of this drive. These dispersive shifts can be turned on and off using a tunable SQUID coupler, meaning a high level of control can be exerted over this system. Such tunable couplers are essential ingredients of today's most performant architectures\cite{Arute2019,Wu:2021od}. 

Previous realizations of a Toffoli gate have had a gate fidelity of $68.5 \%$ with a gate time of $~50$ ns \cite{Fedorov2012ImplementationCircuits} using a technique of hiding the target qubit excitations in higher energy states. Recent realizations of this technique reached $87\%$ process fidelity \cite{Hill2021}, and a very recent implementation of an i-Toffoli  via cross-resonance driving of capacitively coupled transmon qubits reached $98\%$ process fidelity with a gate time of $350$ ns \cite{Kim2021High-fidelityQubits}. 

Our proposal works with transmon qubits coupled via tunable couplers and achieves a process fidelity of over $98\%$ for a gate time of $500$ ns, as we show here via numerical simulations of the process. 
Since we consider the same types of qubits and coupling circuits as implemented in the hardware of most leading developers \cite{Arute2019,Collodo2020ImplementationInteractions,Jurcevic2021DemonstrationSystem,Sete2021FloatingArchitectures,Wu:2021od}, our approach adds further versatility to the possible gates that current quantum chips can run. One of its most important features is that it can be generalized to more than two control qubits, e.g. to CCCX gates, without noticeable loss of fidelity. To this end the frequency of the drive applied to the target qubit needs to be chosen such that it only picks out the specific CCCX transition we require. For current hardware with qubits on a regular 2d grid, this would for example allow to execute not gates controlled by four qubits (as similarly explored with qubits and resonators\cite{Nigg2013StabilizerQubits}). Another generalization of this system is to use more than one drive. This would allow the system to detect qubit cluster parity\cite{Royer2018QubitQED} which is crucial for quantum error correction. 

\begin{figure}[h!]
\includegraphics[width=0.45\textwidth]{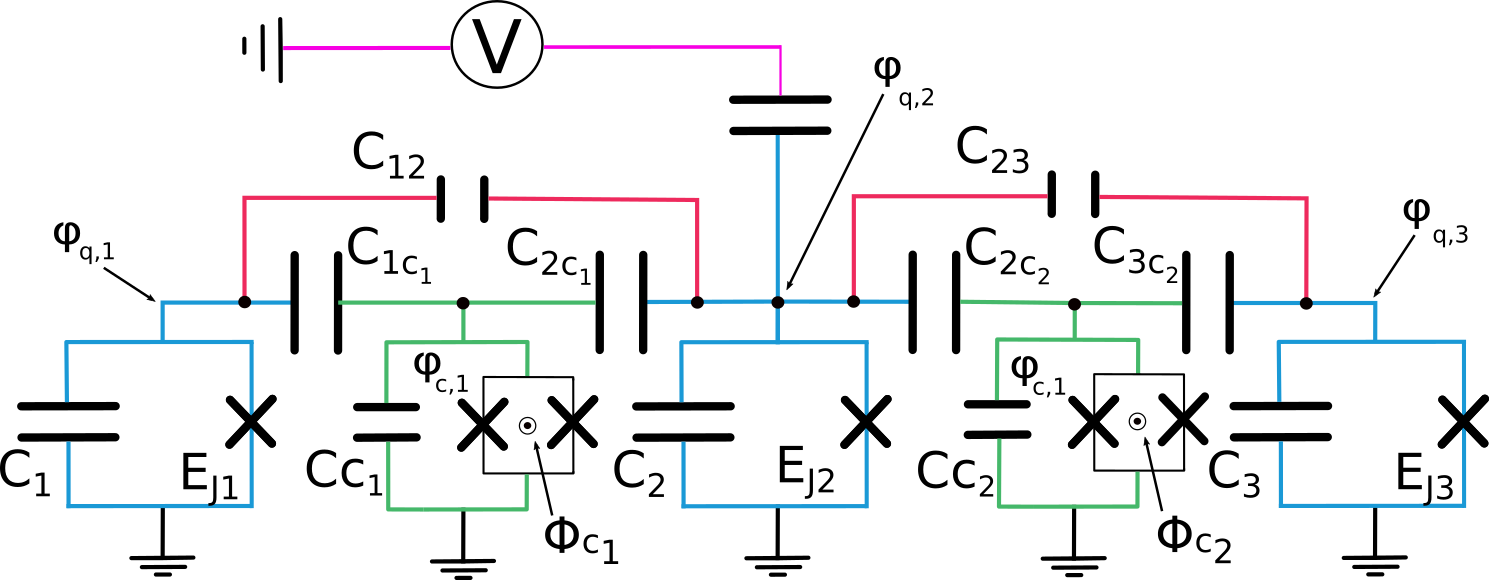}
\caption{Circuit diagram of our system: Here the blue circuits indicate the qubit nodes (described by $\varphi_{q,i}$). Each qubit has a capacitance of $C_i$ and a Josephson Energy of $E_{J,i}$. The green circuits indicate the tunable couplers (described by $\varphi_{c,i} $). The tunable couplers have capacitance $C_{Ci}$, Josephson Energy of $E_{J,cj}$ and they are driven by an external flux $\Phi_{Ci}$ which tunes their frequency. The second qubit is driven by an external voltage (denoted in pink here) which executes the CCX interaction. The qubits are coupled to one another via a small capacitance $C_{ij}$ (red circuits) and the qubits and tunable couplers are each coupled via a capacitance $C_{i,cj}$.}
\label{fig:Circuit} 
\end{figure}

We consider three qubits that are coupled via two tunable couplers, where the qubits are capacitively coupled to the tunable couplers, providing a tunable interaction\cite{Sameti2019FloquetModel,Jin2013PhotonCavities} between the target and control qubits as demonstrated in \cite{Chen2014QubitCoupling} \cite{Yan2018TunableGates,Collodo2020ImplementationInteractions,Li2020TunableCircuit}, see Fig. \ref{fig:Circuit} for a circuit diagram.

By adjusting the coupler transition frequencies with external fluxes we can tune the couplings \cite{Chow2011SimpleQubits,Majer2007CouplingBus} giving versatile control over the interactions of this system. We use transmon qubits \cite{Koch2007} and capacitively shunted dc-SQUIDs as  tunable couplers. After quantization of the circuit and dropping counter rotating terms (see Appendix \ref{Appendix:Derivation} for full derivation) we obtain the Hamiltonian
\begin{align}
    \nonumber
    H &= \sum_{i =1}^3 \left(\omega_i \qd_i q_i + \frac{\alpha_i}{2} \qd_i \qd_i q_i q_i\right)
    + \sum_{j =1}^2 \left(\omega_{cj} \cd_j c_j + \frac{\alpha_{cj}}{2} \cd_j \cd_j c_j c_j\right)
    \\ 
    &- \sum_{n < m=1}^3 g_{nm} (\qd_n q_m +\qd_m q_n)
    - \sum_{k=1}^2 g_{k,c1} (\cd_{1} q_{k} + \qd_{k} c_{1} )
    \\ \nonumber
    &-\sum_{l=2}^3 g_{l,c2} (\cd_{2} q_{l} + \qd_{l} c_{2})
    + \Omega (t) (q_2 + q_2^{\dag}).
\end{align}
where $q_i$ ($c_i$) represent the annihilation operators for the qubits (couplers) and $\omega_i$/$\alpha_i$ ($\omega_{ci}$/$\alpha_{ci}$) are the qubit (coupler) transition frequencies and anharmonicities respectively. The external voltage drive applied to the target qubit is denoted by $\Omega (t)$, $g_{nm}$ denotes the coupling between different qubits (the exact form can be found in Appendix \ref{Appendix:Derivation}) and $g_{i,cj}$ describes the coupling between the i-th qubit and the j-th coupler. We choose capacitances such that the coupling between the qubits is much smaller than the qubit-coupler coupling.
We detune the coupler and qubits by $> 1 $ GHz to ensure the counter rotating terms do not contribute\cite{Yan2018TunableGates}.
In this dispersive regime (i.e. $\frac{g_{i,cj}}{\omega_i - \omega_{cj}} \ll 1)$ we can eliminate the coupler using a Schrieffer-Wolff (SW) transformation \cite{Bravyi2011Schrieffer-WolffSystems}, $H \to \tilde{H} = e^{i S} H e^{-i S}$, where $S = \sum_{i=1}^2 \frac{g_{i,c1}}{\omega_i - \omega_{c1}} (\qd_i c_1 - q_i \cd_1) + \sum_{j=2}^3 \frac{g_{j,c2}}{\omega_j - \omega_{c2}} (\qd_j c_2 - q_j \cd_2)$. Keeping terms up to second order in this expansion we can decouple the qubits from the couplers, such that we are only left with qubit-qubit couplings described by the Hamiltonian,
\begin{equation}\label{eqt:SW_transform_1}
    \tilde{H} =  H_{0,q} + H_{qq} + H_d,
\end{equation}
where ($\hbar = 1$)
\begin{align}\label{eqt:Hamltonian_coupler_slim_mainText}
    H_{0,q} &= \sum_{j =1}^3 \tilde{\omega}_j \qd_j q_j + \frac{\tilde{\alpha}_j}{2} \qd_j \qd_j q_j q_j,
    \\ \nonumber
    H_{qq} &= \sum_{n<m=1}^3 \tilde{g}_{nm}(q_n \qd_m + \qd_n q_m),
    \\ \nonumber
    H_d &= \Omega(t) (q_2 + \qd_2).
\end{align}
Here $\tilde{\omega}_{n}$, $\tilde{\alpha}_{n}$ and $\tilde{g}_{nm}$ are shifted frequencies, nonlinearities and couplings (see Appendix \ref{Appendix:Derivation} for explicit expressions). We have dropped the counter rotating terms and assumed that the coupler always remains in the ground state. The latter allows us to drop the terms describing the coupler as it is no longer coupled to the qubits.

Following the procedure outlined in \cite{Zhu2013CircuitRegime} we use perturbation theory to calculate the corrections to the eigenenergies of $H_{0,q}$ due to the interaction term $V = H_{qq}$. We use these to calculate the dispersive shifts $\chi_{nm}$, where $n,m$ denote the qubits that the shift applies to,
\begin{eqnarray}\label{eqt:Dispersive_shifts_defs_2}
    \chi_{12} &=& E_{\ket{110}} - E_{\ket{100}} - E_{\ket{010}} + E_{\ket{000}},
    \\ \nonumber
    \chi_{13} &=& E_{\ket{101}} - E_{\ket{100}} - E_{\ket{001}} + E_{\ket{000}}.
    \\ \nonumber
     \chi_{23} &=& E_{\ket{011}} - E_{\ket{010}} - E_{\ket{001}} + E_{\ket{000}}.
\end{eqnarray}
and the total shift on the $\ket{111}$ state,
\begin{eqnarray}\label{eqt:Dispersive_shifts_defs_3}
    \chi_{123} &=& E_{\ket{111}} - E_{\ket{100}} - E_{\ket{010}} - E_{\ket{001}} + 2E_{\ket{000}}.
\end{eqnarray}
up to second order. Here $E_{\ket{n}}$ denotes the energy of the state $\ket{n}$ including corrections up to 2nd order in $H_{qq}$. We note that $\chi_{123}$ contains all possible shifts on the $\ket{111}$ state, this will include both controlled phase (CPhase) and controlled controlled phase (CCPhase) shifts (see Appendix \ref{Appendix:Perturbations} for discussion of CCPhase), which emerge because of the finite nonlinearity of transmon qubits, see Eq.(\ref{eqt:dispersive_shift}). The dispersive shifts can be expanded in orders of perturbation theory $\chi^{(n)}$,
\begin{eqnarray}
\chi_\nu = \chi^{(1)}_\nu + \chi^{(2)}_\nu  + ....
\end{eqnarray}
where $\nu \in \{ 12,13,23,123 \}$. We note that all first order terms vanish, $\chi^{(1)}_\nu = 0$, as the interaction term $V$ is off-diagonal. To second order we obtain the following shifts, 
\begin{eqnarray}\label{eqt:dispersive_shift}
    \chi_{ij}^{(2)} &=& 
    -\frac{2\,{\tilde{g}_{ij}}^2}{\tilde{\alpha}_{i}+\tilde{\Delta}_{ij}}
    -\frac{2\,{\tilde{g}_{ij}}^2}{\tilde{\alpha}_{j}+\tilde{\Delta}_{ji}} 
    \\ \nonumber
    &+& \frac{2\,{\tilde{g}_{ij}}^2}{\tilde{\alpha}_{i}+\tilde{\Sigma}_{ij}}+\frac{2\,{\tilde{g}_{ij}}^2}{\tilde{\alpha}_{j}+\tilde{\Sigma}_{ij}} - \frac{4\,{\tilde{g}_{ij}}^2}{\tilde{\alpha}_{i}+\tilde{\alpha}_{j}+\tilde{\Sigma}_{ij}},\\
    \chi_{123}^{(2)} &=& \chi_{12}^{(2)} + \chi_{23}^{(2)} + \chi_{13}^{(2)}, \nonumber
\end{eqnarray}
where we have introduced the notation $\tilde{\Delta}_{ij} = \tilde{\omega}_i - \tilde{\omega}_j$ and $\tilde{\Sigma}_{ij} = \tilde{\omega}_i + \tilde{\omega}_j$.
In a frame, where each qubit rotates at its transition frequency (including second order perturbative corrections) and neglecting $\chi_{13}$ and all shifts of third order and higher by choosing parameters such that these perturbations are highly supressed. We thus arrive at the Hamiltonian
\begin{eqnarray} \label{eq:htilde}
    \tilde{H}_{2-Lvl} &=& 
    \chi_{12}^{(2)} (\ket{110}\bra{110} +\ket{111}\bra{111})
    \\\nonumber
    &+& \chi_{23}^{(2)} (\ket{011}\bra{011} + \ket{111}\bra{111})
    \\ \nonumber
    &+&  H_d,
\end{eqnarray}
in a two level approximation.

\begin{figure*}[h!]
     \centering
    \begin{subfigure}[t]{0.5\textwidth}
    \includegraphics[width=0.8\textwidth]{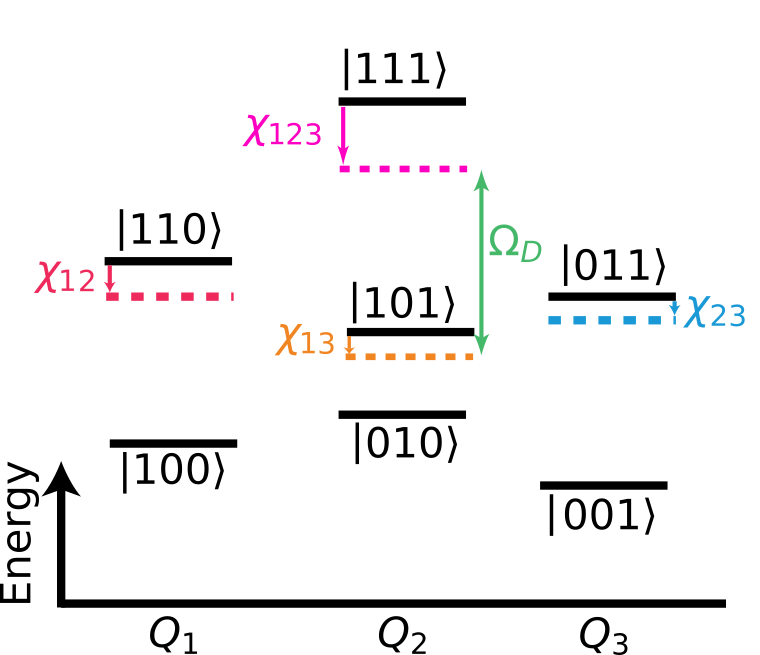}
    \subcaption{ }
    \end{subfigure}%
    ~ 
    \begin{subfigure}[t]{0.5\textwidth}
        \centering
        \includegraphics[width=0.7\textwidth]{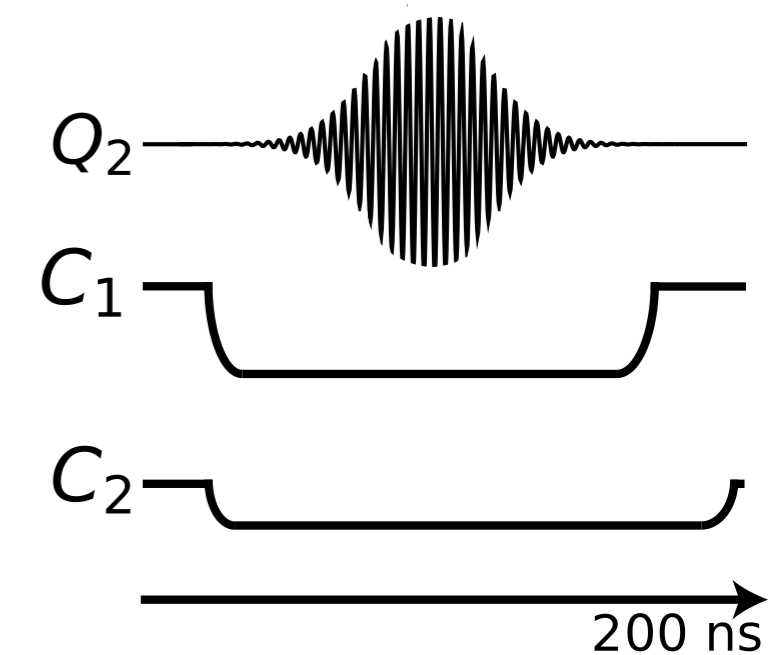}
        \subcaption{ }
    \end{subfigure}
    \caption{{\bf (a)} Energy level diagram detailing the dispersive shifts and couplings within the system. Also showing the applied drive. {\bf (b)} Suggested pulse schemes for an experimental realization. $C_1$ and $C_2$ show the biases applied to the couplers bringing them to the needed frequency to cause the dispersive shifts and $Q_2$ showing the drive pulse applied to qubit 2.}
    \label{fig:Toffoli_Pulse_Shape}
\end{figure*}

The dispersive shifts can be thought of as shifts in the qubit transition frequency dependant on the state of the other qubits. For suitable drive frequencies, this allows us to individually address specific transitions of the three qubit system as each excited qubit adds a contribution to the dispersive shift of the target qubit. This results in the $\ket{110}, \ket{011} $ and $\ket{111}$ states being shifted by $\chi_{12}$, $\chi_{23}$ and $\chi_{12} + \chi_{23}$ respectively. In our case we wish to address the $\ket{101} \leftrightarrow \ket{111}$ transition which will be shifted by $\chi_{12} + \chi_{23}$ by the external drive to produce an i-Toffoli gate. The level diagram and suitable control pulses are sketched in Fig. \ref{fig:Toffoli_Pulse_Shape}

Our approach relies on $ZZ$ interactions that cause the dispersive shifts as an essential ingredient. These dispersive shifts induce conditional phase accumulation on specific states,
which can be described by the unitary $ U_{phase} = \text{diag}[1,1,1,e^{i \chi_{23}  t},1,1,e^{i \chi_{12} t},e^{i (\chi_{12} + \chi_{23}) t}]$.
These conditional phases correspond to CPhase gates that are included in our three qubit gate. One can either work with this generalized Toffoli gate or apply simple strategies to eliminate the contributions of conditional phase gates. 
We can cancel the accumulated phase ($\chi_{ij}$) by applying CPhase gates after the Toffoli gate has been executed. After having corrected for the phases discussed above and removing any single qubit phases we are left with a factor of $-i$ multiplying the flipped states ($\ket{111}$ and $\ket{101}$) \cite{Rasmussen2020Single-stepGates}. If needed, this factor could be removed via the inclusion of an ancilla qubit \cite{Rasmussen2020Single-stepGates}.
We want to add that Toffoli gates with modified phases have been found to be useful in certain quantum algorithms \cite{Cleve1996SchumachersComputation}.

Given the above discussion, we can determine the unitary that we obtain if all employed approximations work perfectly. We take into consideration the accumulated phases discussed above along with the factor of $-i$ caused by the drive only being resonant with the subspace $\{\ket{101},\ket{111}\}$, which we are thus not able to compensate for with a virtual Z gate. Denoting  $\tilde{U}$ the unitary that results from a perturbation free evolution as generated by $\tilde{H}_{2-Lvl}$ as in Eq. (\ref{eq:htilde}) and using $U_{phase}^{\dag}$ as a perfect phase correcting unitary which corrects for the accumulated phases, we find
\begin{equation}
    U_{i-Toffoli} = U_{Phase}^{\dag} \, \tilde{U},
\end{equation}
where $U_{i-Toffoli}$ is defined in Eq. (\ref{eqt:UITOFF}).

We numerically simulated the dynamics generated by the Hamiltonian $\tilde{H}$ of Eq.(\ref{eqt:SW_transform_1}) using QUTIP \cite{Johansson2013} and the q-optimize package \cite{Wittler2021IntegratedQubits}. By performing sweeps over realistic parameter ranges we identified suitable parameters for the circuit and then used the q-optimize package to optimise the drive pulse. We applied DRAG (Derivative Removal by Adiabatic Gate) \cite{Motzoi2009SimpleQubits} to shape the Gaussian pulses such that states outside the computational subspace are not excited despite the presence of the dispersive shifts. 


Denoting by $ U_{sim}$the unitary that results from the simulated dynamics generated by $\tilde{H}$, we quantify the fidelity of the gate by comparing $ U_{phase}^{\dag} U_{sim}$,  and $U_{i-Toffoli}$.
In terms of process fidelity, as measured by the entanglement fidelity $F_p(U_1,U_2) = |\text{Tr} (U_1^\dag , U_2)|/d$ ($d=8$ being the dimension of the Hilbert space), our scheme reaches $F_p(U_{i-Toffoli},U_{phase}^{\dag} U_{sim}) \gtrsim 98 \%$ for the parameters stated below. 

\begin{figure*}[ht!]
    \centering
    \begin{subfigure}[t]{0.5\textwidth}
    \includegraphics[width=1.1\textwidth]{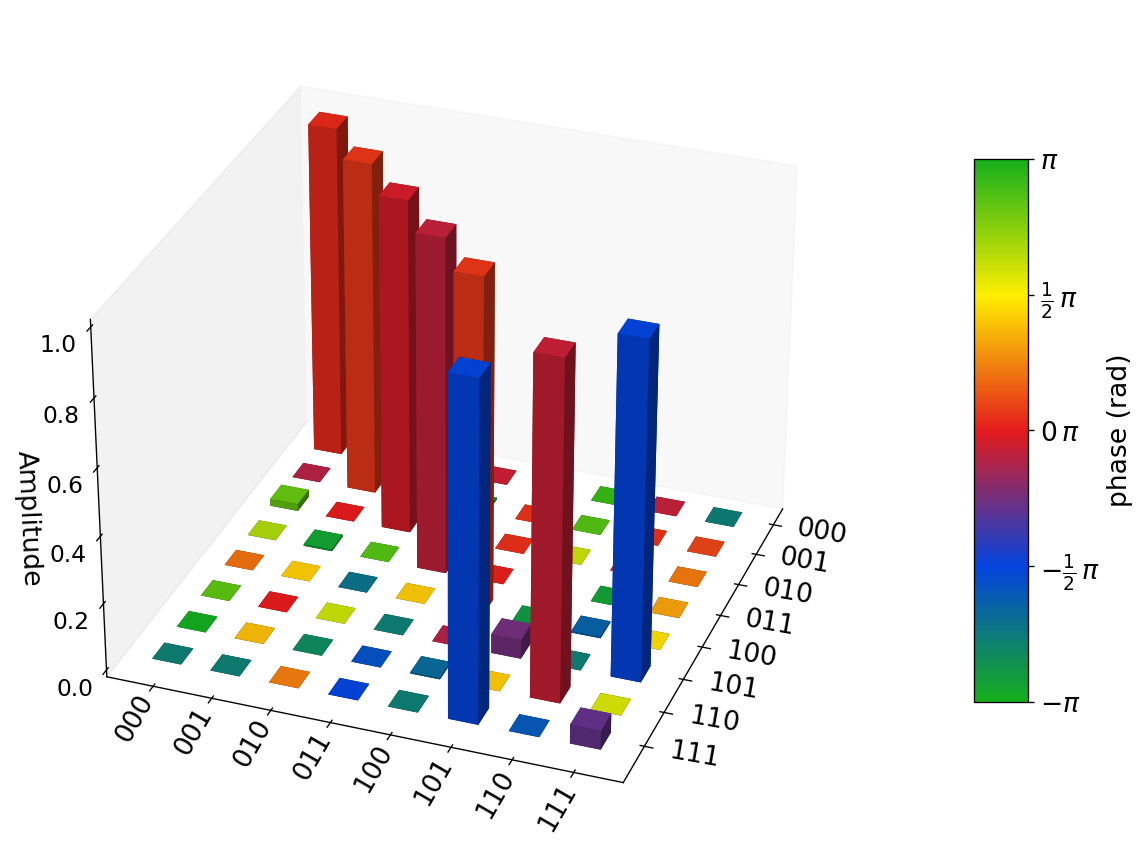}
    \subcaption{ }
    \end{subfigure}%
    ~ 
    \begin{subfigure}[t]{0.5\textwidth}
        \centering
        \includegraphics[width=0.7\textwidth]{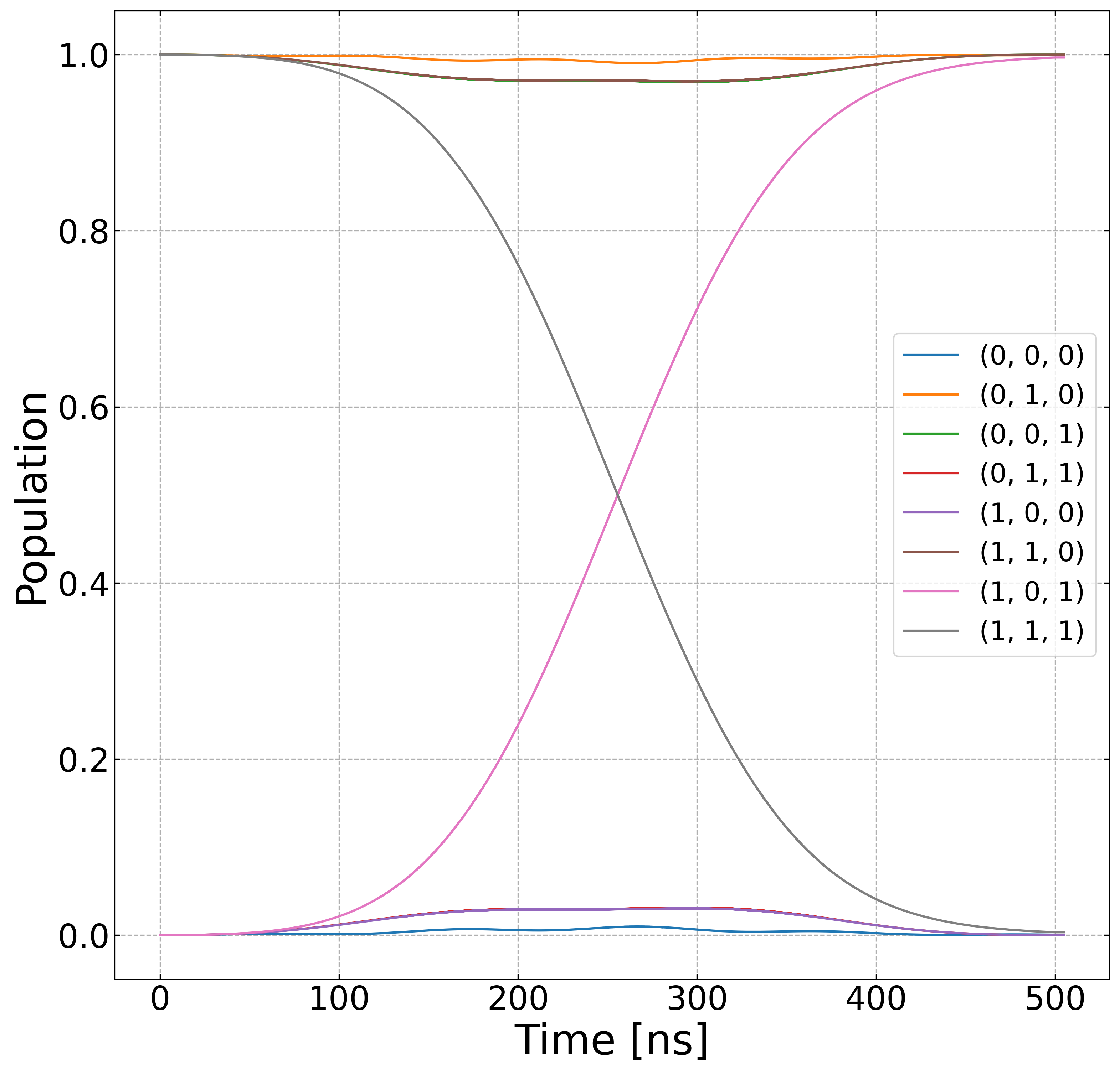}
        \subcaption{ }
    \end{subfigure}
    \caption{{\bf (a)} Absolute values of the matrix elements of $U_{Phase}^{\dag} U_{sim}$ for our scheme, showing $\gtrsim98\%$ process fidelity. Here the evolved unitary has been multiplied by a correcting phase unitary to simulate a perfect phase correction process. As a consequence, all phases have returned close to 0. One sees, that the states $\ket{101}$ and $\ket{111}$ are fully swapped, with an added phase of $-\pi/2$, making the gate an i-Toffoli gate. 
    The system parameters (in natural frequency units) used are: qubit frequencies of $\tilde{\omega}_1 / 2 \pi  = 4.984$ GHz, $\tilde{\omega}_2 / 2 \pi = 5.300$ GHz and $\tilde{\omega}_3 / 2 \pi  = 4.820$ GHz. Anharmonicities of $\tilde{\alpha}_1 / 2 \pi  = \tilde{\alpha}_3 / 2 \pi = -330$ MHz , $\tilde{\alpha} / 2 \pi = -240$ MHz and qubit qubit coupling strengths of $\tilde{g}_{12} / 2 \pi = 15.4$ MHz, $\tilde{g}_{23} / 2 \pi = 29.2$ MHz, $\tilde{g}_{13}/ 2 \pi = 2$ MHz. The peak drive amplitude was $| \Omega | / 2 \pi = 1.5 $ MHz. The phases of this simulation were corrected against the idle phases simulated using the same parameters.  {\bf (b)} Population transfer between all states, the legend on the right hand side depicts the initial state.}
    \label{fig:C3_UnitaryAndPopulation}
\end{figure*}

We choose qubit frequencies $\tilde{\omega}_1 / 2 \pi  = 4.984$ GHz, $\tilde{\omega}_2/ 2 \pi  = 5.300$ GHz, $\tilde{\omega}_3/ 2 \pi  = 4.820$ GHz, so as to maximise fidelity and ensure that our approximations are highly accurate (leading terms of the neglected perturbations are sufficiently small). The peak drive amplitude of the considered Gaussian pulse was $| \Omega | / 2 \pi = 1.5 $ MHz. The corresponding simulation results are shown in Fig. \ref{fig:C3_UnitaryAndPopulation}, see the figure caption for the remaining parameters. 

Alternatively, the gate can also be executed in 350ns with a process fidelity of $\sim 97 \%$ for slightly modified parameters given by; qubit frequencies of $\tilde{\omega}_1 / 2 \pi  = 5.00$ GHz, $\tilde{\omega}_2 / 2 \pi = 5.300$ GHz and $\tilde{\omega}_3 / 2 \pi  = 4.820$ GHz. Anharmonicities of $\tilde{\alpha}_1 / 2 \pi  = \tilde{\alpha}_3 / 2 \pi = -300$ MHz , $\tilde{\alpha} / 2 \pi = -200$ MHz and qubit qubit coupling strengths of $\tilde{g}_{12} / 2 \pi = 19.4$ MHz, $\tilde{g}_{23} / 2 \pi = 35.0$ MHz, $\tilde{g}_{13}/ 2 \pi = 2$ MHz. The peak drive amplitude was $| \Omega | / 2 \pi = 2.5 $ MHz.

The conditions, which limit the achievable process fidelity in our scheme are: (i)
The gate needs to be executed much quicker than the decoherence time of the qubits, where a quicker gate requires a larger drive. (ii) To avoid excitations of higher energy states, the drive needs to remain in the weak driving regime, 
where increasing the drive amplitude ($|\Omega (t)|$) would require a similar increase in the dispersive shifts $\chi_{12} $ and $ \chi_{23}$. The latter could be achieved by decreasing the detuning $\Delta$. (iii) The detuning $\Delta$ however needs to remain large enough to keep the device in the dispersive regime. (iv) The pulse $\Omega(t)$ needs to be such that non-computational states remain unoccupied. (v) The parameters need to be chosen such that the unwanted interactions leading to CCPhase action on the $\ket{111}$ state are minimised. 
The approach to a single shot Toffoli gate that we present here can be modified and enhanced in multiple ways. Firstly, it can also be implemented in a circuit without tunable couplers, e.g. where frequency tunable qubits are  capacitively coupled. The absence of tunable couplers would mean a stronger coupling between the qubits, and since the dispersive shift is proportional to the square of this coupling strength, it would lead to a larger shift in qubit transition frequencies, allowing for larger drives and thus even faster gate times.

For the simple circuit discussed here, the roles of target and control qubits are not interchangeable. Yet for qubits arranged in a triangle, with tunable couplers between each pair of neighbors, each qubit can take the role of the target qubit. Importantly, in a lattice, where each qubit has four direct neighbors, as required for surface code realizations, any two of the four neighboring qubits can take the role of control qubits.
This could also be achieved by connecting all the qubits to a central tunable coupler\cite{McKay2016UniversalBus,Sameti2017SuperconductingCode}, this would allow for tunability of all the couplings via modification of the coupler frequency.

Further extensions of this system can be envisaged by adding more qubits to the circuit thus allowing us to execute higher order Toffoli gates (Multi-Controlled-Not-Gates), such as for example a CCCX gate. As with the standard Toffoli gate, higher order Toffoli gates can be decomposed into single- and two-qubit gates. The number of elementary gates in such expansions however scales exponentially with the number of control qubits involved \cite{Nielsen2000QuantumInformation}, and in some cases can require many more ancilla qubits \cite{Barenco1995ElementaryComputation}. The realization of single-shot higher order controlled NOT gates following our scheme would circumvent this problem and allow the use of these gates without modification of existing hardware. These higher order gates have uses in quantum information algorithms \cite{Shor1995SchemeMemory}, quantum error correction \cite{Shor1995SchemeMemory,Steane1996ErrorTheory,Nigg2013StabilizerQubits} and quantum annealing \cite{Chancellor2017CircuitArchitecture}. 
Since current decomposition's of higher order gates require $2^{n-1}$ Controlled gates (where n is the number of control qubits) \cite{Barenco1995ElementaryComputation}, the fidelity of such decomposed operations would be significantly lower than the single shot implementation we propose here.


In summary we have presented a proposal for a single step Toffoli gate using dispersive shifts. We have numerically simulated the system and shown that the achievable process fidelity and gate time are significantly better than most current implementations  and comparable to latest results within superconducting circuits. The proposed implementation uses existing superconducting devices and is thus straight-forward to implement in available hardware. The approach we present generalizes in a straight forward manner to higher order controlled gates such as CCCX gates which are useful in quantum error correction, where parity measurements using this method could also be executed. These higher order gates could also be useful for quantum simulators of high energy physics or quantum chemistry simulators where three-body interactions are crucial for emulating interactions between gauge and matter fields.

\begin{acknowledgements}
This work has received funding from the European Union’s Horizon 2020 research and innovation program under grant agreement No 828826 “Quromorphic", and the MSCA Cofund action No 847471 "Qustec", from the German Federal Ministry of Education and Research via the funding program quantum technologies - from basic research to the market under contract number 13N15684 and 13N15680 "GeQCoS", and support from EPSRC DTP grant EP/R513040/1.
\end{acknowledgements}

\newpage
\appendix
\clearpage
\newpage

\section{Derivation of Circuit Hamiltonian}\label{Appendix:Derivation}


We begin the derivation with the Lagrangian of the circuit in figure \ref{fig:Circuit} where we have defined $C_i (C_{ci})$ to be the capacitance of the qubits (couplers), $C_{icj} (C_{ij})$ to be the qubit-coupler (qubit-qubit) couplings and $E_{J,qi} (\tilde{E}_{J,cj})$ are the Josephson Energies associated with the qubits (couplers).
\begin{eqnarray}
    L_{circuit} & = &\sum_{i = 1} ^3 \phizerotwo C_{q,i} \Dot{\varphi}_{q,i}^2 + E_{J,qi} \cos(\vari{q,i}) 
    \\ \nonumber
    &+& \sum_{j = 1}^2 \phizerotwo C_{cj} \Dot{\varphi}_{c,j}^2 + \tilde{E}_{J,cj}(\varphi_{e,j}) \cos(\vari{c,j})
    \\ \nonumber
    & -&  \phizerotwo \sum_{k=1}^2 C_{kc1} (\Dot{\varphi}_k - \Dot{\varphi}_{c1})^2 - \phizerotwo \sum_{l=2}^3 C_{lc2} (\Dot{\varphi}_l - \Dot{\varphi}_{c2})^2 
    \\ \nonumber
    &-&  \phizerotwo C_{12} (\Dot{\varphi}_1 - \Dot{\varphi}_2)^2 - \phizerotwo C_{23} (\Dot{\varphi}_2 - \Dot{\varphi}_3)^2.
\end{eqnarray}
Where $\vari{i}$ is the phase associated with the nodes indicated in figure \ref{fig:Circuit}, $\Phi_0 = \frac{\hbar}{2 e}$ and $2 E_{J,cj} \cos(\frac{\vari{e,j}}{2}) \equiv \tilde{E}_{J,cj}(\varphi_{e,j})$ where $\vari{e,j} = \frac{\Phi_e}{\phizero}$ and $\Phi_e$ is the external flux. Computing the Legendre transform we obtain the Hamiltonian of the circuit
\begin{eqnarray}
    H_{circuit} &= &\sum_{i =1}^3 \frac{\pi_{qi}^2}{2  \phizero \Bar{C}_{q,i}} - E_{J,qi} \cos(\vari{qi}) 
    +\sum_{j=1}^2 \frac{\pi_{cj}^2}{2  \phizero \Bar{C}_{c,j}}
    \\ \nonumber
    &-& \tilde{E}_{J,cj}(\varphi_{e,j}) \cos(\vari{cj})
    - \sum_{l=1}^2 \frac{\pi_l \pi_{c1} }{2  \Bar{C}_{l,c1} \phizero} 
    \\ \nonumber
    &-& \sum_{k=2}^3 \frac{\pi_k \pi_{c2} }{2  \Bar{C}_{k,c2} \phizero} 
    -\sum_{n < m=1}^3 \frac{\pi_{qn} \pi_{qm}}{2  \Bar{C}_{n,m} \phizero} .
\end{eqnarray}
with $\pi_X$ being the conjugate momenta to $\varphi_X$ and the following definitions for the capacitances
\begin{eqnarray}
    \Bar{C}_{q1} &=& C_{q1} + C_{12} + C_{1,c1},
    \quad \Bar{C}_{q3} = C_{q3} + C_{23} + C_{3,c2},
    \\ \nonumber
    \Bar{C}_{q2} &=& C_{q2} + C_{12} + C_{23} + C_{2,c1} + C_{2,c2},
    \\ \nonumber
    \Bar{C}_{c1} &=& C_{c1} + C_{1,c1} + C_{2,c1}, 
    \\ \nonumber
    \Bar{C}_{c2} &=& C_{c2} + C_{2,c2} + C_{3,c2},
    \\ \nonumber
    \frac{1}{\Bar{C}_{12}} &=& \frac{-2 C_{12}}{\Bar{C}_{q1} \Bar{C}_{q2}},
    \quad
    \frac{1}{\Bar{C}_{23}} = \frac{-2 C_{23}}{\Bar{C}_{q2} \Bar{C}_{q3}},
    \quad
    \frac{1}{\Bar{C}_{13}} = \frac{4 C_{12} C_{23}}{\Bar{C}_{q1} \Bar{C}_{q2} \Bar{C}_{q3}},
    \\ \nonumber
    \frac{1}{\Bar{C}_{1,c1}} &=& \frac{4 C_{12} C_{2,c1} - 2 C_{1,c1} \Bar{C}_{q2}}{\Bar{C}_{q1} \Bar{C}_{q2} \Bar{C}_{c1}} \approx \frac{-2 C_{1,c1}}{\Bar{C}_{q1} \Bar{C}_{c1}},
    \\ \nonumber
    \frac{1}{\Bar{C}_{1,c2}} &=& \frac{4 C_{12} C_{2,c2} \Bar{C}_{q3} - 8 C_{12} C_{23} C_{3,c2}}{\Bar{C}_{q1} \Bar{C}_{q2} \Bar{C}_{q3} \Bar{C}_{c2}} \approx \frac{4 C_{12} C_{2,c2} }{\Bar{C}_{q1} \Bar{C}_{q2} \Bar{C}_{c2}},
    \\ \nonumber
    \frac{1}{\Bar{C}_{2,c1}} &=& \frac{-2 \Bar{C}_{2,c1}}{\Bar{C}_{q2} \Bar{C}_{c1}},
    \quad \frac{1}{\Bar{C}_{2,c2}} = \frac{4 C_{23} C_{3,c2} - 2 C_{2,c2} \Bar{C}_{q3}}{\Bar{C}_{q2} \Bar{C}_{q3} \Bar{C}_{2c}} \approx \frac{-2 C_{2,c2}}{\Bar{C}_{q2} \Bar{C}_{c2}},
    \\ \nonumber
    \frac{1}{\Bar{C}_{3,c2}} &=& \frac{-2 \Bar{C}_{3,c2}}{\Bar{C}_{q3} \Bar{C}_{c2}}.
\end{eqnarray}
where we choose the capacitances such that they obey a 
hierarchy i.e. $C_{i,j} \ll C_{k,cl} \ll C_{m}$, this allows us to make the approximations above. We also define a general charging energy to be
\begin{equation}
    E_{CX} = \frac{e^2}{2 C_{X}}.
\end{equation}
With these definitions in place we quantize the Hamiltonian by defining creation and annihilation operators via $\vari{i} = \Tilde{\varphi}_i (q_i - \qd_i)$ and $\pi_i = \frac{i \hbar}{2 \Tilde{\vari{i}}} (q_i - \qd_i)$ with canonical commutation relations, $[q_i,\qd_j] = \delta_{ij}$ and similar relations for the coupler variables, $\vari{ci} = \Tilde{\varphi}_{ci} (c_i - \cd_i)$ and $\pi_{ci} = \frac{i \hbar}{2 \Tilde{\vari{ci}}} (c_i - \cd_i)$ with $[c_i,\cd_j] = \delta_{ij}$ and $[c_i,\qd_j] = [\cd_i,q_j] =  0$. Here using these operators and setting $\hbar = 1$ we get,
\begin{eqnarray}
\label{eqt:Bare_Hamiltonian}
    H &=& \sum_{i =1}^3 \omega_{qi} \qd_i q_i + \frac{\alpha_{qi}}{2} \qd_i \qd_i q_i q_i
    \\ \nonumber
    &+& \sum_{j =1}^2 \omega_{cj} \cd_j c_j + \frac{\alpha_{cj}}{2} \cd_j \cd_j c_j c_j
    \\ \nonumber
    &+&\sum_{n < m =1}^3 g_{nm} (q_n - \qd_n)(q_m - \qd_m)
    - \sum_{k=1}^2 g_{k,c1}(q_k c_1^{\dag} +  q_k^{\dag} c_1)
    \\ \nonumber
    &-&\sum_{k'=2}^3 g_{k',c2}(q_{k'} c_2^{\dag} +  q_{k'}^{\dag} c_2)
    + \Omega (t) (q_2 + q_2^{\dag}).
\end{eqnarray}
Where we have dropped the counter rotating terms in the qubit-qubit and qubit-coupler terms. In this Hamiltonian we have defined the following parameters
\begin{eqnarray}
    \omega_{qi} &=& \sqrt{8 E_{Cqi} E_{J,qi}} - E_{Cqi},
    \quad
    \alpha_{qi} = -E_{Cqi},
    \\ \nonumber
    \omega_{ci} &=& \sqrt{8 E_{Ci} E_{J,ci}} - E_{Ci},
    \quad
    \alpha_{ci} = -E_{Ci},
    \\ \nonumber
    \Tilde{\varphi}_{qi} &=& \left(\frac{2 E_{Cqi}}{E_{J,qi}} \right)^{\frac{1}{4}}, \quad \Tilde{\varphi}_{ci} = \left(\frac{2 E_{Ci}}{E_{J,ci}} \right)^{\frac{1}{4}},
    \\ \nonumber
    g_{ij} &=& \frac{1}{2} \frac{\Bar{C}_{ij}}{\sqrt{\Bar{C}_{qi} \Bar{C}_{qj}}} \sqrt{\omega_{qi} \omega_{qj}},
    \quad 
    g_{j,ci} = \frac{1}{2} \frac{\Bar{C}_{j,ci}}{\sqrt{\Bar{C}_{qj} \Bar{C}_{ci}}} \sqrt{\omega_{qj} \omega_{ci}}.
\end{eqnarray}

\begin{table}[h!]
\caption{\label{tab:paramsSimulation}Suggested parameters for circuit and the calculated dispersive shifts that result from these parameters.}
    \begin{ruledtabular}
        \begin{tabular}{c c c}
            Element &$\omega_{X}/ (2 \pi) $ [GHz]& $\alpha_i / 2 \pi$ [MHz]\\
            \hline 
            Qubit 1 & 4.99 & -300\\
            Qubit 2 & 5.31 & -250 \\
            Qubit 3 & 4.83 & -300\\
            Coupler 1 & 7 & -200 \\
            Coupler 2 & 6.8 & -200 
        \end{tabular}
    \end{ruledtabular}
\end{table}

\begin{table}[h!]
\centering
    \begin{ruledtabular}
        \begin{tabular}{c c} 
            \multicolumn{2}{c}{Couplings [MHz]}\\
            \hline 
            $g_{12}/ 2 \pi$   &12  \\
            $g_{13}/ 2 \pi$   &2  \\
            $g_{23}/ 2 \pi$   &10.5  \\
            $g_{1,c1}/ 2 \pi$ &55 \\
            $g_{2,c1}/ 2 \pi$ &55 \\
            $g_{2,c2}/ 2 \pi$ &130 \\
            $g_{3,c2}/ 2 \pi$ &130 
        \end{tabular}
    \end{ruledtabular}
\end{table}

\begin{table}[h!]
\centering
    \begin{ruledtabular}
        \begin{tabular}{c c} 
            \multicolumn{2}{c}{Dispersive Shifts [MHz]}\\
            \hline 
            $\chi_{12} / 2 \pi$ & -5.1 \\
            $\chi_{23}/ 2 \pi$ & -4.95 \\
            $\chi_{13}/ 2 \pi$ & 0.04 \\
            $\chi_{123}/ 2 \pi$ & 0.63 \\
        \end{tabular}
    \end{ruledtabular}
\end{table}
With the quantized Hamiltonian in place we turn to deriving the dispersive shifts mentioned in the main text. We perform a Schrieffer-Wolff (SW) transformation $ \Tilde{H} = e^{-S} H e^S$ to eliminate the coupling to the SQUID with $S = \sum_{i=1}^2\frac{g_{i,c1}}{\omega_{qi} - \omega_{c1}} (\qd_i c_1 - q_i \cd_1) + \sum_{j=2}^3\frac{g_{j,c2}}{\omega_{qj} - \omega_{c2}} (\qd_j c_2 - q_j \cd_2)$. This transformation is valid under the conditions $\frac{g_{j,c1}}{\Delta_{j,c1}} \ll 1$ and $\frac{g_{j,c2}}{\Delta_{j,c2}} \ll 1$ where $\Delta_{ij} = \omega_i - \omega_j$. We are left with the Hamiltonian
\begin{eqnarray}\label{eqt:Hamiltonian_coupler_eliminated}
    \Tilde{H} &=& \sum_{i=1}^3 \tilde{\omega}_{qi} \qd_i q_i + \frac{\tilde{\alpha}_i}{2} \qd_i \qd_i q_i q_i
    \\ \nonumber
    &+& \sum_{n < m=1}^3 \tilde{g}_{nm}(q_n \qd_m + \qd_n q_m)
     \\ \nonumber
     &+& H_{0,c} + H_d .
\end{eqnarray}
Where $H_{0,c}$ is the free Hamiltonian for the couplers and we have added the term $H_d$ that describes the applied drive. Here we have expanded in the bare coupling which is of the form $g_{icj}^{k1} g_{nm}^{k2}$ and have kept $g_{nm}$ to linear order (as it is considered a second order small quantity) and $g_{icj}$ to second order. The new coupling strengths and transition frequencies are defined as
\begin{eqnarray} \label{eqt:Dressed_variables}
\tilde{\omega}_{q1} &=& \omega_{q1} +  \frac{g_{1,c1}^2}{\Delta_{1,c1}} , \quad \tilde{\omega}_{q3} = \omega_{q3} +  \frac{g_{3,c2}^2}{\Delta_{3,c2}},
\\ \nonumber
\tilde{\omega}_{q2} &=& \omega_{q2} +  \frac{g_{2,c1}^2}{\Delta_{2,c1}} + \frac{g_{2,c2}^2}{\Delta_{2,c2}},
\\ \nonumber
\tilde{g}_{12} &=&g_{12} +  g_{1,c1} g_{2,c1}\left(\frac{1}{\Delta_{1,c1}} + \frac{1}{\Delta_{2,c1}} \right), 
\\ \nonumber
\tilde{g}_{23} &=& g_{23} + g_{2,c2} g_{3,c2}\left(\frac{1}{\Delta_{2,c2}} + \frac{1}{\Delta_{3,c2}} \right),
\\ \nonumber
\tilde{g}_{13} &=&g_{13}.
\end{eqnarray}

\section{Perturbations }\label{Appendix:Perturbations}
\subsection{Schrieffer–Wolff Approximation}
At higher order in the SW approximation there are hopping terms between the couplers and qubits. Whilst the couplers should stay in their ground during the gate, we still estimate the strength of the interaction between the couplers and qubits to check if it may cause problems. The strengths of these interactions are given by
\begin{eqnarray}
  \tilde{g}_{1,c1} = \frac{g_{1,2} g_{1,c1}}{\Delta_{1,c1}}
  \quad \text{, }
  \tilde{g}_{2,c1} = \frac{g_{1,2} g_{2,c1}}{\Delta_{2,c1}} 
  \\ \nonumber
  \tilde{g}_{2,c2} = \frac{g_{2,3} g_{2,c2}}{\Delta_{2,c2}} 
  \quad \text{, }
  \tilde{g}_{3,c2} = \frac{g_{2,3} g_{3,c2}}{\Delta_{3,c2}},
\end{eqnarray}
we can calculate these strengths using the parameters mentioned above, to be $\leq 1 $MHz. However since the qubits are far detuned from the couplers these interaction should not contribute to the dynamics of the system.
\\
\section{Higher order Accumulated Phase}
Here we estimate the dispersive shift $\chi_{13}$ and third order dispersive shifts, that have been neglected in our approach.
\subsection{Dispersive Shift between Qubit 1 and 3}
In Eq.(\ref{eqt:Hamiltonian_coupler_eliminated}) there still remains a term proportional to $(q_1 q_3^{\dag} + h.c)$ we assume this is small enough to neglect in the main text. Here we calculate it's effect. This term will contribute a dispersive shift to the states $\ket{101}$ and $\ket{111}$. This should not effect our scheme much as both of these states are shifted by the same amount to second order calculated by
\begin{eqnarray}
    \chi_{13}^{(2)} &=& 
    -\frac{2\,{\tilde{g}_{13}}^2}{\tilde{\alpha}_{1}+\tilde{\Delta}_{13}}
    -\frac{2\,{\tilde{g}_{13}}^2}{\tilde{\alpha}_{3}+\tilde{\Delta}_{13}} 
    \\ \nonumber
    &+& \frac{2\,{\tilde{g}_{13}}^2}{\tilde{\alpha}_{1}+\tilde{\Sigma}_{13}}+\frac{2\,{\tilde{g}_{13}}^2}{\tilde{\alpha}_{3}+\tilde{\Sigma}_{13}} - \frac{4\,{\tilde{g}_{13}}^2}{\tilde{\alpha}_{1}+\tilde{\alpha}_{3}+\tilde{\Sigma}_{13}}.
    \end{eqnarray}
This amounts to a shift of $\approx 0.1$ MHz for our parameters.
\subsection{Third Order Correction to Dispersive Shifts}
We truncated our two body dispersive shifts to second order. Higher order terms still remain, and for exact calculations they must be used, see \cite{Sung2021RealizationCoupler} for a detailed calculation of the third order corrections. For three body terms we calculated the shifts up to third order in perturbation theory the full expression for $\chi_{123}^{(3)}$ is given below, for our parameter we find $\chi_{123}^{(3)}/ 2 \pi \approx 0.6$ MHz. This CCPhase that will be accumulated on the $\ket{111}$ state cannot be corrected for via two-qubit gates thus will have to be eliminated via aligning parameters so the gate time and the period of $\chi_{123}^{(3)}$ align at a multiple of $2 \pi$. Alternatively one can operate in a regime where $\chi_{123}^{(3)}$ is very small and thus can be safely ignored.
\begin{widetext}
\begin{eqnarray} \label{eqt:chi123(3)}
\chi_{123}^{(3)} &=& 
\\ \nonumber
2 \tilde{g}_{12} \tilde{g}_{23} \tilde{g}_{13} &\bigg(&\frac{1}{\tilde{\Delta}_{12} \tilde{\Delta}_{13}}+\frac{1}{\tilde{\Sigma}_{12} \tilde{\Delta}_{13}}-\frac{2}{(\tilde{\alpha}_{1}+\tilde{\Delta}_{12})
(\tilde{\alpha}_{1}+\tilde{\Delta}_{13})}-\frac{4}{(\tilde{\alpha}_{1}+\tilde{\alpha}_{2}+\tilde{\Sigma}_{12}) (\tilde{\alpha}_{1}+\tilde{\Delta}_{13})}-\frac{1}{\tilde{\Delta}_{12}
\tilde{\Delta}_{23}}+\frac{1}{\tilde{\Sigma}_{12} \tilde{\Delta}_{23}}+\frac{1}{\tilde{\Delta}_{13} \tilde{\Delta}_{23}}
\\\nonumber
&-&\frac{2}{(\tilde{\alpha}_{2}-\tilde{\Sigma}_{12})
(\tilde{\alpha}_{2}+\tilde{\Delta}_{23})}-\frac{4}{(\tilde{\alpha}_{1}+\tilde{\alpha}_{2}+\tilde{\Sigma}_{12}) (\tilde{\alpha}_{2}+\tilde{\Delta}_{23})}-\frac{4}{(\tilde{\alpha}_{1}+\tilde{\Delta}_{13})
(\tilde{\alpha}_{2}+\tilde{\Delta}_{23})}-\frac{4}{(\tilde{\alpha}_{2}-\tilde{\Sigma}_{12}) (\tilde{\alpha}_{3}
-\tilde{\Sigma}_{13})}
\\\nonumber
&+&\frac{2}{\tilde{\Sigma}_{12} (\tilde{\alpha}_{3}-\tilde{\Sigma}_{13})}+\frac{1}{\tilde{\Delta}_{12}
\tilde{\Sigma}_{13}}+\frac{2}{(\tilde{\alpha}_{2}-\tilde{\Sigma}_{12}) \tilde{\Sigma}_{13}}-\frac{3}{\tilde{\Sigma}_{12} \tilde{\Sigma}_{13}}+\frac{2}{(\tilde{\alpha}_{2}+\tilde{\Sigma}_{12})
\tilde{\Sigma}_{13}}-\frac{1}{\tilde{\Delta}_{23} \tilde{\Sigma}_{13}}+\frac{2}{(\tilde{\alpha}_{2}+\tilde{\Delta}_{23}) \tilde{\Sigma}_{13}}
\\\nonumber
&+&\frac{2}{(\tilde{\alpha}_{1}+\tilde{\Sigma}_{12})
(\tilde{\alpha}_{1}+\tilde{\Sigma}_{13})}+\frac{2}{\tilde{\Sigma}_{12} (\tilde{\alpha}_{3}+\tilde{\Sigma}_{13})}-\frac{4}{(\tilde{\alpha}_{1}+\tilde{\Delta}_{12}) (\tilde{\alpha}_{1}+\tilde{\alpha}_{3}+\tilde{\Sigma}_{13})}-\frac{8}{(\tilde{\alpha}_{1}+\tilde{\alpha}_{2}+\tilde{\Sigma}_{12})
(\tilde{\alpha}_{1}+\tilde{\alpha}_{3}+\tilde{\Sigma}_{13})}
\\\nonumber
&-&\frac{4}{(\tilde{\alpha}_{1}+\tilde{\Delta}_{12}) (\tilde{\alpha}_{3}-\tilde{\Sigma}_{23})}+\frac{2}{\tilde{\Sigma}_{12}
(\tilde{\alpha}_{3}-\tilde{\Sigma}_{23})}-\frac{2}{(\tilde{\alpha}_{3}-\tilde{\Sigma}_{13}) (\tilde{\alpha}_{3}-\tilde{\Sigma}_{23})}-\frac{4}{(\tilde{\alpha}_{1}+\tilde{\alpha}_{3}+\tilde{\Sigma}_{13})
(\tilde{\alpha}_{3}-\tilde{\Sigma}_{23})}-\frac{1}{\tilde{\Delta}_{12} \tilde{\Sigma}_{23}}
\\\nonumber
&+&\frac{2}{(\tilde{\alpha}_{1}+\tilde{\Delta}_{12}) \tilde{\Sigma}_{23}}-\frac{3}{\tilde{\Sigma}_{12}
\tilde{\Sigma}_{23}}+\frac{2}{(\tilde{\alpha}_{1}+\tilde{\Sigma}_{12}) \tilde{\Sigma}_{23}}-\frac{1}{\tilde{\Delta}_{13} \tilde{\Sigma}_{23}}+\frac{2}{(\tilde{\alpha}_{1}+\tilde{\Delta}_{13})
\tilde{\Sigma}_{23}}-\frac{3}{\tilde{\Sigma}_{13} \tilde{\Sigma}_{23}}+\frac{2}{(\tilde{\alpha}_{1}+\tilde{\Sigma}_{13}) \tilde{\Sigma}_{23}}
\\\nonumber
&+&\frac{2}{(\tilde{\alpha}_{2}+\tilde{\Sigma}_{12})
(\tilde{\alpha}_{2}+\tilde{\Sigma}_{23})}+\frac{2}{\tilde{\Sigma}_{13} (\tilde{\alpha}_{2}+\tilde{\Sigma}_{23})}+\frac{2}{\tilde{\Sigma}_{12} (\tilde{\alpha}_{3}+\tilde{\Sigma}_{23})}+\frac{2}{(\tilde{\alpha}_{3}+\tilde{\Sigma}_{13})
(\tilde{\alpha}_{3}+\tilde{\Sigma}_{23})}
\\\nonumber
&-&\frac{4}{(\tilde{\alpha}_{2}-\tilde{\Sigma}_{12}) (\tilde{\alpha}_{2}+\tilde{\alpha}_{3}+\tilde{\Sigma}_{23})}-\frac{8}{(\tilde{\alpha}_{1}+\tilde{\alpha}_{2}+\tilde{\Sigma}_{12})
(\tilde{\alpha}_{2}+\tilde{\alpha}_{3}+\tilde{\Sigma}_{23})}-\frac{4}{(\tilde{\alpha}_{3}-\tilde{\Sigma}_{13}) (\tilde{\alpha}_{2}+\tilde{\alpha}_{3}+\tilde{\Sigma}_{23})}
\\\nonumber
&-&\frac{8}{(\tilde{\alpha}_{1}+\tilde{\alpha}_{3}+\tilde{\Sigma}_{13})
(\tilde{\alpha}_{2}+\tilde{\alpha}_{3}+\tilde{\Sigma}_{23})}\bigg)
\end{eqnarray}
\end{widetext}

\subsection{Third Order Correction to two qubit dispersive shifts}
These are expressed by a large number of terms so for the sake of clarity we only write the corrections to the energies
\begin{widetext}
\begin{eqnarray}
\nonumber
E_{\ket{110}}^{(3)} = \tilde{g}_{12} \tilde{g}_{23} \tilde{g}_{13} &\bigg(&
\frac{4}{\tilde{\Delta}_{13}\,\left(\tilde{\alpha}_{2}-\tilde{\Sigma}_{12}\right)}+\frac{4}{\tilde{\Delta}_{23}\,\left(\tilde{\alpha}_{1}+\tilde{\Delta}_{12}\right)}-\frac{2}{\tilde{\Sigma}_{12}\,\tilde{\Delta}_{13}}-\frac{2}{\tilde{\Sigma}_{12}\,\tilde{\Delta}_{23}}-\frac{8}{\left(\tilde{\alpha}_{1}+\tilde{\Sigma}_{13}\right)\,\left(\tilde{\alpha}_{2}+\tilde{\Sigma}_{23}\right)}+\frac{4}{\tilde{\Delta}_{23}\,\left(\tilde{\alpha}_{1}+\tilde{\Sigma}_{13}\right)}
\\\nonumber
&+&\frac{4}{\tilde{\Delta}_{13}\,\left(\tilde{\alpha}_{2}+\tilde{\Sigma}_{23}\right)}-\frac{2}{\tilde{\Delta}_{13}\,\tilde{\Delta}_{23}}-\frac{8}{\left(\tilde{\alpha}_{1}+\tilde{\Sigma}_{13}\right)\,\left(\tilde{\alpha}_{1}+\tilde{\alpha}_{2}+\tilde{\Sigma}_{12}\right)}-\frac{8}{\left(\tilde{\alpha}_{2}+\tilde{\Sigma}_{23}\right)\,\left(\tilde{\alpha}_{1}+\tilde{\alpha}_{2}+\tilde{\Sigma}_{12}\right)}
\\
&-&\frac{4}{\left(\tilde{\alpha}_{1}+\tilde{\Delta}_{12}\right)\,\left(\tilde{\alpha}_{1}+\tilde{\Sigma}_{13}\right)}-\frac{4}{\left(\tilde{\alpha}_{2}-\tilde{\Sigma}_{12}\right)\,\left(\tilde{\alpha}_{2}+\tilde{\Sigma}_{23}\right)}\bigg)
\\\nonumber
E_{\ket{101}}^{(3)} =\tilde{g}_{12} \tilde{g}_{23} \tilde{g}_{13} &\bigg(& \frac{4}{\tilde{\Delta}_{12}\,\left(\tilde{\alpha}_{3}-\tilde{\Sigma}_{13}\right)}-\frac{4}{\tilde{\Delta}_{23}\,\left(\tilde{\alpha}_{1}+\tilde{\Delta}_{13}\right)}-\frac{2}{\tilde{\Sigma}_{13}\,\tilde{\Delta}_{12}}+\frac{2}{\tilde{\Sigma}_{13}\,\tilde{\Delta}_{23}}-\frac{8}{\left(\tilde{\alpha}_{1}+\tilde{\Sigma}_{12}\right)\,\left(\tilde{\alpha}_{3}+\tilde{\Sigma}_{23}\right)}-\frac{4}{\tilde{\Delta}_{23}\,\left(\tilde{\alpha}_{1}+\tilde{\Sigma}_{12}\right)}
\\\nonumber
&+&\frac{4}{\tilde{\Delta}_{12}\,\left(\tilde{\alpha}_{3}+\tilde{\Sigma}_{23}\right)}+\frac{2}{\tilde{\Delta}_{12}\,\tilde{\Delta}_{23}}-\frac{8}{\left(\tilde{\alpha}_{1}+\tilde{\Sigma}_{12}\right)\,\left(\tilde{\alpha}_{1}+\tilde{\alpha}_{3}+\tilde{\Sigma}_{13}\right)}-\frac{8}{\left(\tilde{\alpha}_{3}+\tilde{\Sigma}_{23}\right)\,\left(\tilde{\alpha}_{1}+\tilde{\alpha}_{3}+\tilde{\Sigma}_{13}\right)}
\\
&-&\frac{4}{\left(\tilde{\alpha}_{1}+\tilde{\Delta}_{13}\right)\,\left(\tilde{\alpha}_{1}+\tilde{\Sigma}_{12}\right)}-\frac{4}{\left(\tilde{\alpha}_{3}-\tilde{\Sigma}_{13}\right)\,\left(\tilde{\alpha}_{3}+\tilde{\Sigma}_{23}\right)}\bigg)
\\\nonumber
E_{\ket{011}}^{(3)} =\tilde{g}_{12} \tilde{g}_{23} \tilde{g}_{13} &\bigg(& \frac{2}{\tilde{\Sigma}_{23}\,\tilde{\Delta}_{12}}-\frac{4}{\tilde{\Delta}_{13}\,\left(\tilde{\alpha}_{2}+\tilde{\Delta}_{23}\right)}-\frac{4}{\tilde{\Delta}_{12}\,\left(\tilde{\alpha}_{3}-\tilde{\Sigma}_{23}\right)}+\frac{2}{\tilde{\Sigma}_{23}\,\tilde{\Delta}_{13}}-\frac{8}{\left(\tilde{\alpha}_{2}+\tilde{\Sigma}_{12}\right)\,\left(\tilde{\alpha}_{3}+\tilde{\Sigma}_{13}\right)}-\frac{4}{\tilde{\Delta}_{13}\,\left(\tilde{\alpha}_{2}+\tilde{\Sigma}_{12}\right)}
\\\nonumber
&-&\frac{4}{\tilde{\Delta}_{12}\,\left(\tilde{\alpha}_{3}+\tilde{\Sigma}_{13}\right)}-\frac{2}{\tilde{\Delta}_{12}\,\tilde{\Delta}_{13}}-\frac{8}{\left(\tilde{\alpha}_{2}+\tilde{\Sigma}_{12}\right)\,\left(\tilde{\alpha}_{2}+\tilde{\alpha}_{3}+\tilde{\Sigma}_{23}\right)}-\frac{8}{\left(\tilde{\alpha}_{3}+\tilde{\Sigma}_{13}\right)\,\left(\tilde{\alpha}_{2}+\tilde{\alpha}_{3}+\tilde{\Sigma}_{23}\right)}
\\
&-&\frac{4}{\left(\tilde{\alpha}_{2}+\tilde{\Delta}_{23}\right)\,\left(\tilde{\alpha}_{2}+\tilde{\Sigma}_{12}\right)}-\frac{4}{\left(\tilde{\alpha}_{3}-\tilde{\Sigma}_{23}\right)\,\left(\tilde{\alpha}_{3}+\tilde{\Sigma}_{13}\right)}\bigg)
\\
E_{\ket{100}}^{(3)} =\tilde{g}_{12} \tilde{g}_{23} \tilde{g}_{13} &\bigg(& \frac{2}{\tilde{\Sigma}_{23}\,\tilde{\Delta}_{12}}-\frac{4}{\tilde{\Sigma}_{23}\,\left(\tilde{\alpha}_{1}+\tilde{\Sigma}_{13}\right)}-\frac{4}{\tilde{\Sigma}_{23}\,\left(\tilde{\alpha}_{1}+\tilde{\Sigma}_{12}\right)}+\frac{2}{\tilde{\Sigma}_{23}\,\tilde{\Delta}_{13}}-\frac{4}{\left(\tilde{\alpha}_{1}+\tilde{\Sigma}_{12}\right)\,\left(\tilde{\alpha}_{1}+\tilde{\Sigma}_{13}\right)}-\frac{2}{\tilde{\Delta}_{12}\,\tilde{\Delta}_{13}}\bigg)
\\
E_{\ket{010}}^{(3)} =\tilde{g}_{12} \tilde{g}_{23} \tilde{g}_{13} &\bigg(& \frac{2}{\tilde{\Sigma}_{13}\,\tilde{\Delta}_{23}}-\frac{4}{\tilde{\Sigma}_{13}\,\left(\tilde{\alpha}_{2}+\tilde{\Sigma}_{23}\right)}-\frac{2}{\tilde{\Sigma}_{13}\,\tilde{\Delta}_{12}}-\frac{4}{\tilde{\Sigma}_{13}\,\left(\tilde{\alpha}_{2}+\tilde{\Sigma}_{12}\right)}-\frac{4}{\left(\tilde{\alpha}_{2}+\tilde{\Sigma}_{12}\right)\,\left(\tilde{\alpha}_{2}+\tilde{\Sigma}_{23}\right)}+\frac{2}{\tilde{\Delta}_{12}\,\tilde{\Delta}_{23}}\bigg)
\\
E_{\ket{001}}^{(3)} =\tilde{g}_{12} \tilde{g}_{23} \tilde{g}_{13} &\bigg(& -\frac{4}{\tilde{\Sigma}_{12}\,\left(\tilde{\alpha}_{3}+\tilde{\Sigma}_{13}\right)}-\frac{4}{\tilde{\Sigma}_{12}\,\left(\tilde{\alpha}_{3}+\tilde{\Sigma}_{23}\right)}-\frac{2}{\tilde{\Sigma}_{12}\,\tilde{\Delta}_{13}}-\frac{2}{\tilde{\Sigma}_{12}\,\tilde{\Delta}_{23}}-\frac{4}{\left(\tilde{\alpha}_{3}+\tilde{\Sigma}_{13}\right)\,\left(\tilde{\alpha}_{3}+\tilde{\Sigma}_{23}\right)}-\frac{2}{\tilde{\Delta}_{13}\,\tilde{\Delta}_{23}}\bigg)
\\
E_{\ket{000}}^{(3)} =\tilde{g}_{12} \tilde{g}_{23} \tilde{g}_{13} &\bigg(& -\frac{2}{\tilde{\Sigma}_{12}\,\tilde{\Sigma}_{13}}-\frac{2}{\tilde{\Sigma}_{12}\,\tilde{\Sigma}_{23}}-\frac{2}{\tilde{\Sigma}_{13}\,\tilde{\Sigma}_{23}}\bigg)
\end{eqnarray}
\end{widetext}
\clearpage
\newpage
\bibliography{References}
\end{document}